\documentclass{iau} 
\usepackage{graphicx}

\title{Intra-cluster GC-LMXB in the Fornax galaxy cluster} 

\author{G.Riccio$^1$, M.Paolillo$^{1,2,3}$, R.D'Abrusco$^4$, M.Cantiello$^5$, X.Jin$^6$, Z.Li$^6$, A.Venhola$^7$, G.D'Ago$^8$, T.Puzia$^8$, E.Iodice$^2$, V.Pota$^2$, N.Napolitano$^2$, M.Hilker$^9$, S.Mieske$^{10}$, M.Spavone$^2$, R.Peletier$^{11}$, P.Eigenthaler$^8$}   

\affiliation{$^1$Department of Physics, University of Napoli ''Federico II'', Italy - $^2$INFN - Sezione di Napoli, 80126 Napoli, Italy - $^3$INAF-Oss.di Capodimonte, 80131 Napoli - $^4$Harvard-Smithsonian Center for Astrophysics, Cambridge, MA 02138, USA - $^5$INAF-Oss.dell'Abruzzo, 64100 Teramo, Italy - $^6$School of Astronomy and Space Science, Nanjing University, China - $^7$Astronomy Research Unit, University of Oulu, Finland - $^8$Pontificia Universidad Cat\`olica de Chile, Santiago, Chile - $^9$European Southern Observatory, Garching bei München, Germany - $^{10}$European Southern Observatory, Santiago, Chile - $^{11}$Kapteyn Institute, University of Groningen, the Netherlands}

\pubyear{2019}
\volume{351}  
\jname{Star Clusters: From the Milky Way to the Early Universe}
\editors{A. Bragaglia, M.B. Davies, A. Sills \& E. Vesperini, eds.}
\begin{document}

\maketitle

\begin{abstract}
The formation of Low mass X-ray binaries (LMXB) is favored within dense stellar systems such as Globular Clusters (GCs). The connection between LMXB and Globular Clusters has been extensively studied in the literature, but these studies have always been restricted to the innermost regions of galaxies. We present a study of LMXB in GCs within the central 1.5 $deg^2$ of the Fornax cluster with the aim of confirming the existence of a population of LMXB in intra-cluster GCs and understand if their properties are related to the host GCs, to the environment or/and to different formation channels.
\keywords{X-rays: binaries, galaxies: star clusters, galaxies: clusters: individual (Fornax)}
\end{abstract}

\firstsection

\section{Introduction}
Low mass X-ray binaries are stellar binary systems composed of an extremely dense object (a neutron star or a black hole) accreting mass from the secondary star (1$M_{\odot}$), emitting mainly in UV and X-ray bands. The formation of LMXB in GCs is influenced by several properties including host galaxy type, GC mass, size, central concentration and metallicity. In fact several studies have shown that in early type galaxies the fraction of LMXB residing in GCs varies from 10\%-20\% in small galaxies reaching $\sim 70$\% in cD galaxies, depending on the morphological type of galaxy and on the specific abundance of GCs (e.g. \cite[Kim et al. 2009]{Kim_etal09}). It was also observed that LMXBs tend to form in bright GCs as expected if the luminosity is a proxy to the total number of stars they contain; on the other hand size and concentration reflect the efficiency of dynamical interaction and favor binary formation in dense environments. Moreover red (metal-rich) GCs are $\sim$3 times more likely to host LMXB than blue (metal-poor) GCs (\cite[Jordan et al. 2004]{Jordan_etal04}, \cite[Fabbiano 2006]{Fabbiano06}, \cite[Paolillo et al. 2011]{Paolillo_etal11}, \cite[D'Ago et al. 2014]{D'Ago_etal14} and references therein). 

An important observational problem which limits our knowledge of the factors that drive the formation of LMXBs in GCs is the fact that most studies are limited to the central regions of galaxies, due to the limited area surveyed by past surveys.
In this work we try to expand these results to large galactocentric distances, performing a wide-field study of the LMXBs in intra-cluster GCs in the Fornax galaxy cluster, with the aim of understanding i) if there actually exists a population of intra-cluster GC-LMXB and ii) if this population shares the same properties of the galactic one. The Fornax cluster, located at a distance of 19 Mpc, is the second closest cluster of galaxies (after Virgo), and therefore represents an ideal target for this type of study.

\section{Observations}
The optical data used here were acquired as part of the  Fornax Deep Survey (FDS, \cite[Iodice et al. 2016]{Iodice_etal16}) based on observations obtained in u,g,r,i bands with the VST telescope at the ESO Paranal Observatory. The data reduction was performed using the Astro-WISE pipeline (\cite[Venhola et al. 2019]{Venhola_etal19}). Since GCs, at the distance of Fornax, appear unresolved from the ground, they are hard to separate from stars and compact background galaxies and therefore the selection of the GC sample represents a crucial step in our analysis. In this case we adopted the approach presented in \cite[Cantiello et al. 2018]{Cantiello_etal18} applied to the improved FDS dataset (Cantiello et al., in prep., also see Cantiello et al. in this volume); the GC selection criteria was defined on a training set of spectroscopically confirmed GCs published in  \cite[Pota et al. (2018)]{Pota18} and \cite[Schuberth et al. (2010)]{Schuberth_etal10}. The adopted selection criteria are given in table 1.

\begin{figure}[htb]
    \centering
    \includegraphics[width=1.0\textwidth]{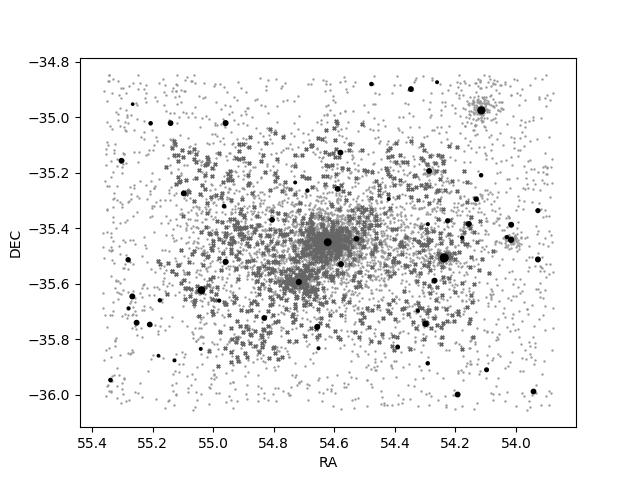}
    \caption{Spatial distribution of the GCs (light grey point) and the X-ray sources (dark grey crosses) centered on NGC1399. Cluster galaxies are the solid black circles with size proportional to their effective radii.}
    \label{fig:d'abrusco}
\end{figure}

 The X-ray data are extracted from archival Chandra observations with a total exposure time of 1.3 Ms covering large part of the Fornax cluster central region. The X ray catalog was extracted by \cite[Jin et al.(2019)]{Jin_etal19} using the Chandra Interactive Analysis of Observation (CIAO) tool wavdetect, finding 1175 X-ray sources in our FoV. Adopting a 1'' matching radius between the X-ray and optical centroids, we identify 168 X-ray sources positionally coincident with GCs. In Figure \ref{fig:d'abrusco} we show the spatial distribution of the objects in our field of view. 
The figure reveals an extended distribution of GCs, as already discovered by \cite[D'Abrusco et al.(2016)]{D'Abrusco_etal16}; while a large fraction of these GCs are clustered around the brightest galaxies, many GCs occupy the intra-cluster space, several effective radii away from any cluster galaxy member.

\begin{table}
  \begin{center}
  \caption{GC candidates selection criteria.}
  \label{tab1}
 {\scriptsize
  \begin{tabular}{|l|c|}\hline 

\textbf{magnitude} & $18\leq m_g \leq 26$  \\ \hline
\textbf{concentration index$^1$} & $0.8 \leq CI_n \leq 1.15$ \\ \hline
\textbf{color} &$0.6 \leq g-i \leq 1.45$ \\
 & $1.35 \leq u-r \leq 3.5$ \\ \hline
\textbf{Difference from model$^2$} & $\leq 0.4$ \\ \hline

  \end{tabular}
  }
 \end{center}
\vspace{1mm}
 \scriptsize{
 {\it Notes:}\\
  $^1$\textit{$CI_n$}: normalized concentration index, based on the difference in \textit{g}-band magnitude between apertures of 6 and 12 pixels. 
  $^2$Maximum distance from the best-fit population synthesis model for spectroscopically confirmed GCs in the color-color diagram}
\end{table}

\begin{figure}[htb]
    \centering
    \includegraphics[width=0.45\textwidth]{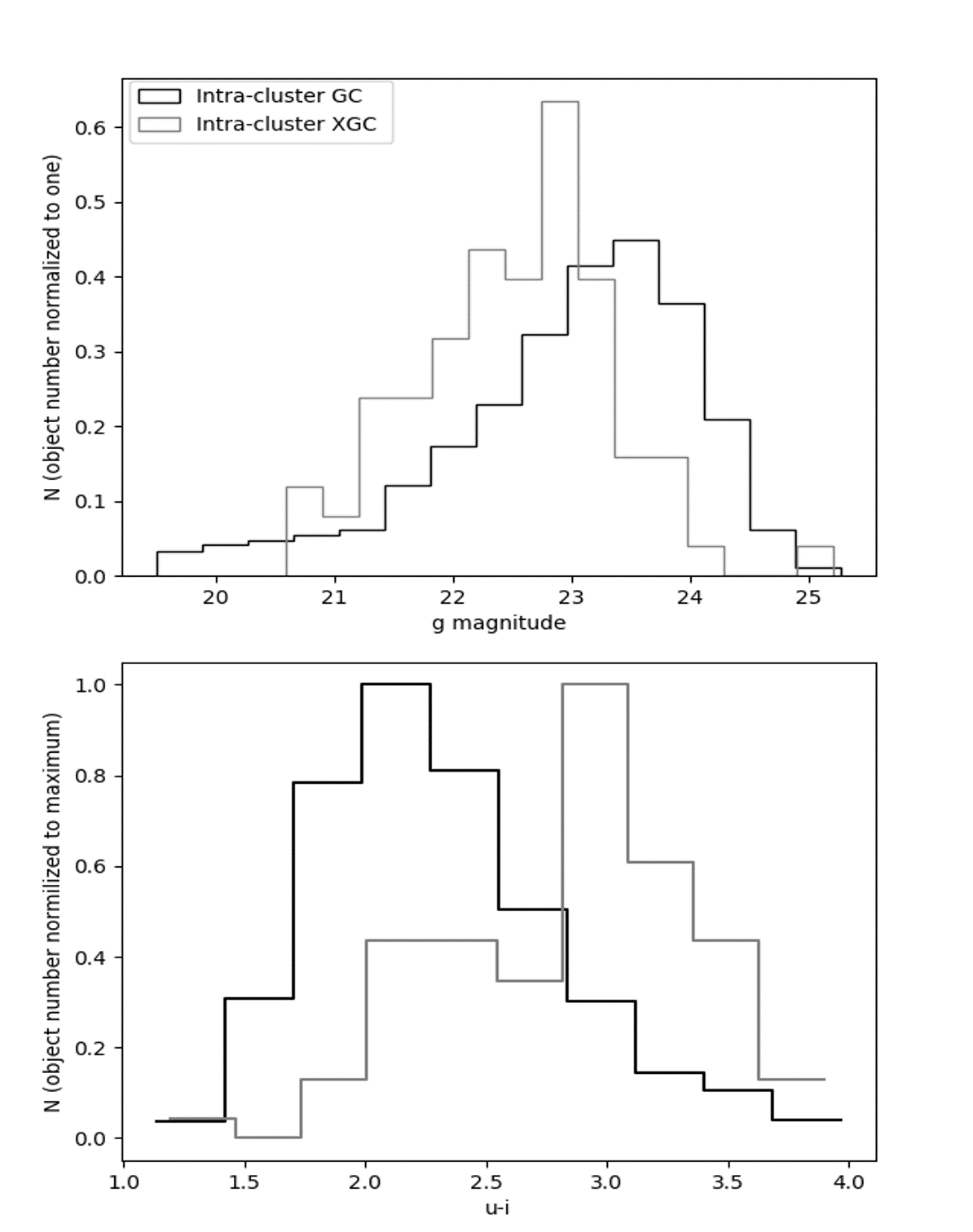}\includegraphics[width=0.6\textwidth]{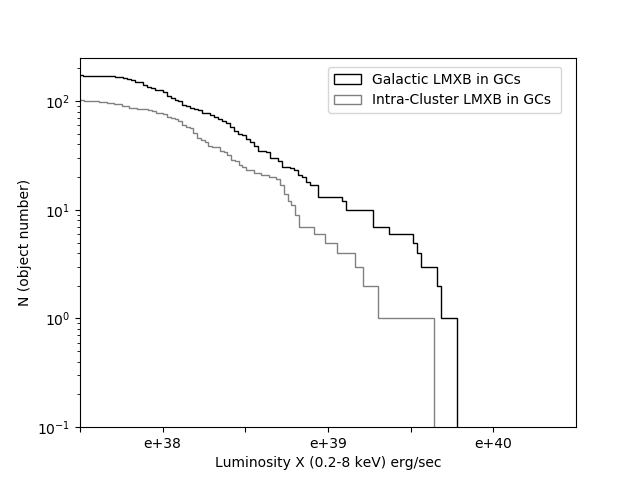}
    \caption{\textit{Left}: magnitude (upper panel) and color (lower panel) distribution of GCs hosting (gray) and not-hosting (black) LMXBs.  \textit{Right:} X-ray luminosity function for galactic and intra-cluster GC-LMXBs.}
    \label{fig:Lf}
\end{figure}

\section{Results}
We analyzed the photometric properties of the GCs that host LMXBs, dividing them into ''galactic'' and ''intra-cluster'' systems based on their distance from the nearest galaxy in terms of effective radii. We consider intra-cluster GCs those lying more than $6 r_e$ from the nearest galaxy, finding 86 intra-cluster GC-LMXB and 82 galactic GC-LMXB. 
Figure \ref{fig:Lf}, left panel shows that intra-cluster LMXBs tend to reside in bright and red GCs, as already found for the galactic population in the past. In fact, despite the fact that the intra-cluster population is dominated by blue GCs, almost the 70\% of LMXB are in red globular clusters.

We analyzed the X-ray properties of the LMXBs associated to intra-cluster GCs. Comparing the X-ray luminosity function of LMXBs in intra-cluster GCs with the galactic ones (Figure \ref{fig:Lf}) we found that the LF of both population follows a power law (down to the completeness limit of our data) with slopes consistent with those found in the past for galactic GC-LMXBs. The slope of the galactic sample is $\alpha=2.03\pm0.12$ while the slope of the intra-cluster one is $\alpha=2.06\pm0.12$. A possible lack of bright LMXBs is observed in intra-cluster systems, but at this stage a Kolmogorov-Smirnov test indicates that this difference is not statistically significant.

\begin{figure}[htb]
    \centering
    \includegraphics[width=1.0\textwidth]{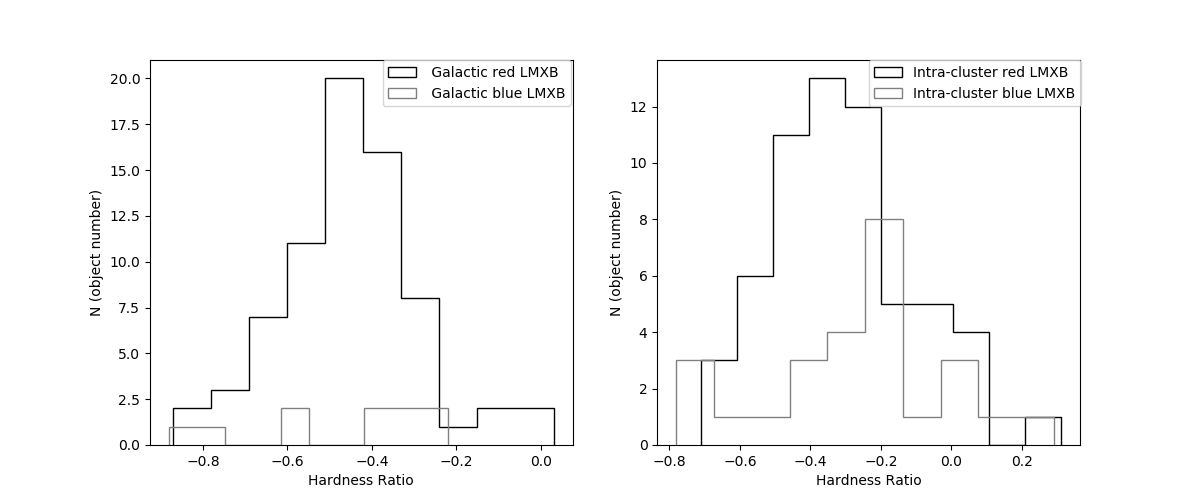}
    \caption{Hardness-ratios of galactic (left) LMXBs and intra-cluster (right) hosted by red and blue GCs. }
    \label{fig:HR}
\end{figure}

An unespected discovery has been the finding of a significant difference in hardness-ratio between the intra-cluster and the galactic GC-LMXB: the intra-cluster sources seem to have harder spectra than the galactic ones. Using a Gaussian Mixture Model we divided the GCs sample in red and blue and we found that the LMXBs in blue GCs tend to be harder than the red ones  (Figure \ref{fig:HR}) which may explain the harder spectrum of intra-cluster sources which are dominated by blue GCs. This result is still tentative and we are evaluating possible systematic effects: a contamination from background AGN seems excluded since we expect only 6-7 random matches; other possible problems could be the contamination due to the diffuse gas in the cluster center, or uncertainties in the spectral response correction due to the combination of the multiple \textit{Chandra} exposures and the off-axis angles. We plan to perform a detailed spectral analysis of the LMXB population in order to confirm and understand the origin of this difference. 

\end{document}